\documentclass{aa}  
\usepackage{graphicx}
\usepackage{natbib}

\usepackage{txfonts}
\begin{document}
   \title{Altair - the 'hottest' magnetically active star in X-rays}

%   \subtitle{}

   \author{J. Robrade
          \and
          J.H.M.M. Schmitt
          }

%   \offprints{J. Robrade}

        \institute{Universit\"at Hamburg, Hamburger Sternwarte, Gojenbergsweg 112, D-21029 Hamburg, Germany\\
       \email{jrobrade@hs.uni-hamburg.de}
             }

   \date{Received 17 November 2008 / Accepted 22 February 2009}

% \abstract{}{}{}{}{} 
% 5 {} token are mandatory
 
  \abstract
  % context heading (optional), leave it empty if necessary  
{The onset of stellar magnetic activity is related to the operation of dynamo processes that require the 
development of an outer convective layer. This transition of stellar interior structure is expected to occur in late A-type stars.}
   {The A7 star Altair is one of the hottest magnetically active stars. Its proximity to the Sun allows a detailed investigation of a corona in X-rays for a star with a shallow convection zone.}
   {We used a deep XMM-Newton observation of Altair and analyzed X-ray light curves, spectra, and emission lines. We investigated the temporal behavior and properties of the
X-ray emitting plasma and studied the global coronal structure of Altair.}
   {Altair's corona with an X-ray luminosity of $L_{\rm X}=1.4\times 10^{27}$~erg/s and an activity level of $\log L_{\rm X}/L_{\rm bol}= -7.4$
is located predominantly at low latitude regions and exhibits X-ray properties that are overall very similar to those of other weakly active stars. 
The X-ray emission is dominated by cool plasma (1\,--\,4~MK) at low density, and elemental abundances exhibit a solar-like FIP effect and Ne/O ratio.
The X-ray brightness varies by 30\,\% over the observation, most likely due to rotational modulation and minor activity;
in contrast, no strong flares or significant amounts of hot plasma were detected. 
The X-ray activity level of Altair is apparently close to the saturation level, which is reduced by roughly four orders of magnitude when compared to late-type stars.}
   {With its fast rotation, Altair provides an inefficient, but very stable dynamo that mainly operates in convective layers below its `cooler' surface areas around the equator. 
This dynamo mechanism results in magnetic activity and leads to X-ray properties that are similar to those of the quiescent Sun, despite very different underlying stars.}

   \maketitle
%
%________________________________________________________________

\section{Introduction}

\object{Altair} ($\alpha$\,Aql, HD~187642, HR~7557, GJ~768) with a spectral type of A7\,IV-V
is a well-studied, optically bright star (V\,=\,0.77, B-V\,=\,0.22) with a mass of 1.8\,M$_{\sun}$ and an age of about 1.2\,Gyr \citep{lac99,sou05}.
It is the earliest and therefore hottest, magnetically active, single late-type main sequence star detected by the {\it Einstein} satellite \citep{gol83,schmitt85} 
and later by ROSAT \citep{schmitt97}. 
Since located  at a distance of only 5.1\,pc, Altair is a unique target for studying the X-ray properties of late A-type stars in greater detail. 

Altair is known to be a very fast rotator from spectroscopic measurements. 
Its rotational velocity has been determined to $Vsini \approx$~190\,--\,250~km/s \citep[e.g.][]{abt95, sto68},
hence Altair rotates at a significant fraction of its breakup-speed, which is estimated to be roughly 400~km/s.
Altair's shape and surface were successively spatially resolved by near-infrared interferometry \citep{bel01,sou05,pet06,mon07}.
The star was found to be oblate, i.e. significantly rotationally deformed, with an axial ratio of a\,/\,b\,$\approx$\,1.1\,--\,1.2, it has
an inclination of the rotational axis to the line of sight of about 60\degr~and an equatorial radius of 1.9\,--\,2.0~R$_{\sun}$.
Modeling of the interferometric data independently confirmed the fast rotation ({$Vsini$\,=\,210\,--\,240~km/s), revealed strong gravity darkening and possibly indicated differential rotation.
The distortion of the stellar photosphere leads to
a latitude-dependent distribution of surface temperatures ranging from around 6900\,K at the equator up to roughly 8500\,K at its poles, while 
classical determinations yielded a range of surface averaged temperature values of $T_{\rm eff}\approx$~7600\,--\,8000~K.
The fast and possible differential rotation, the distortion of the stellar surface, and gravity darkening may all significantly affect the classification of Altair; for example
\cite{pet06} proposes that, instead of being evolved, Altair is a zero-age main sequence star.
Altair was also discovered as a multi-mode pulsating, low-amplitude (mmag) variable star that lies in the $\delta$\,Scuti instability strip \citep{buz05}.

X-ray observations of a large sample of main sequence stars with shallow convection zones - including Altair - performed with {\it Einstein}
indicate that stellar activity begins around late A-type stars and increases strongly towards later, i.e. cooler, stars \citep{schmitt85}.
Actually, Altair's coronal activity level as expressed by the $\log L_{\rm X}/L_{\rm bol}$ ratio has a value of -7.46,
by far the lowest in the sample presented by \cite{schmitt85}, despite its fast rotation.
Observations with ROSAT have confirmed Altair as a weak, soft X-ray source with an X-ray luminosity of $\log L_{\rm X}= 27.4$~erg/s.
In contrast, deep ROSAT PSPC pointings of hotter stars resulted in stringent upper limits for the A0V star Vega at a level of $\log L_{\rm X}\le 25.55$~erg/s and for the
A3V star Fomalhaut at $\log L_{\rm X}\le 25.90$~erg/s. On the other hand, the A7 star Alderamin ($\alpha$\,Cep) at a distance of 15\,pc
is even a RASS (ROSAT All Sky Survey) source with an X-ray luminosity comparable to those of Altair. 
This indicates that coronal X-ray emission from single main sequence stars, at least at nowadays detectable levels, indeed sets in around spectral type A7.
More recently, also {\it Chandra} observations failed to detect X-ray emission from Vega, but yielded one of the lowest upper limits on 
X-ray luminosity ($L_{\rm X}\le 1.8 \times 10^{25}$~erg/s) obtained so far \citep{pea06}.

Besides X-rays, Ly\,$\alpha$ emission that is broadened by $Vsini$\,=220\,km/s has also been detected from Altair by IUE ({\it International Ultraviolet Explorer})
and was identified as a chromospheric component \citep{fer95}. 
The analysis of HST/GHRS spectra including \ion{Si}{iii} and \ion{N}{v} showed the presence of a transition region \citep{sim97}; also  supported by
FUSE ({\it Far Ultraviolet Spectroscopic Explorer}) spectra that cover prominent \ion{C}{iii} and \ion{O}{vi} emission lines. These lines show very broad profiles
and the emission was clearly attributed to Altair itself, confirming the
presence of a transition region and ruling out a `hidden' companion as origin of the far UV and by implication coronal X-ray emission \citep{red02}.
Several attempts have been made to measure the magnetic field of Altair, e.g. by circular polarization measurement that traces the longitudinal component, 
but resulted in its non-detection at an accuracy of about 100\,G \citep{lan82}.

The onset of stellar magnetic activity is connected to the development of an outer convective layer and
its understanding is fundamental for dynamo mechanisms and coronal heating in late-type stars in general.  
Observations with ROSAT have shown the formation of X-ray emitting coronae around late-type cool dwarf 
stars to be universal \citep{schmitt97}, whereas activity levels
span a broad range of values from $\log L_{\rm X}/L_{\rm bol} \approx -7~...-3$. The tight correlation of $L_{\rm X}$ with 
rotation or more precisely dynamo efficiency then suggests a magnetic character of their activity in analogy to the Sun.
In a solar-like dynamo magnetic activity is generated
at the interface layer between radiative core and outer convection zone, the so-called $\alpha\,\Omega$ dynamo. Its efficiency is characterized by the inverse square of
the Rossby number (Ro\,=\,$P/\tau_{c}$), defined as the ratio between rotational period and convective turnover time, whereas
magnetic activity saturates for fast rotators at an activity level of  $\log L_{\rm X}/L_{\rm bol} \approx -3$.
However, towards hotter, more massive stars the outer convection zone becomes increasingly thinner and therefore the dynamo
efficiency declines strongly even for fast rotating stars.
In this scenario Altair's very low activity level points to the presence of a very inefficient dynamo due to a large Rossby number, i.e. a thin convection zone/small turnover time.
In the regime of mid A-type stars at effective temperatures above $T_{\rm eff}\approx 8200$\,K the outer convective layer vanishes, as expected theoretically
and as indicated by FUSE observations of chromospheric emission lines \citep{sim02}.
Consequently, magnetic activity is absent or extremely weak in these stars.

X-ray emission was also detected from several stars of earlier spectral type.
In a comparison study of several hundred ROSAT X-ray sources and positions from the Bright Star Catalog,
the detection fraction is fairly constant at a level around 0.1 over the spectral range B3\,--\,A6 \citep{schroeder07}.
It increases strongly towards early B-stars, the realm of wind driven X-rays from hot stars,
as well as towards early F-stars, the realm of magnetic activity of cool stars.
However, when compared to late A-type stars,
the X-ray detected stars in the spectral range from mid-B to mid-A exhibit harder X-ray spectra and/or significantly higher X-ray luminosities.
Many of these are pre-main sequence Herbig Ae/Be stars or peculiar Ap/Bp stars, where
different intrinsic X-ray generation mechanisms may operate, e.g. a magnetic shear dynamo in young Ae/Be stars \citep{tou95}
or magnetically confined wind-shocks as proposed for the A0p star IQ~Aur \citep{bab97}. Otherwise
unresolved binary components are suspected to be responsible for the observed X-ray emission.
Altogether, a common origin of their X-ray emission with those of Altair appears rather unlikely.

In this paper we present a deep {\it XMM-Newton} observation of Altair.
Beside light curves and medium resolution spectra it provides the first well exposed high resolution X-ray spectrum of a late A-type star so far.
This unique data set allows us to study X-ray variability, plasma properties like temperatures, emission measures and abundances
and to address the topics of coronal structures, Ne/O ratio or dynamo efficiency in a star with a very shallow convection zone.
Our paper is structured as follows; in Sect.\,\ref{ana} we describe the data used and analysis methods, in Sect.\,\ref{ress} we present our results,
discuss our findings in Sect.\,\ref{dis} and in Sect.\,\ref{con} we give our conclusions.

\section{Observations and data analysis}
\label{ana}

Altair was observed by {\it XMM-Newton} during two observing blocks separated by two weeks in October 2007 for a total of roughly 130\,ks.
{\it XMM-Newton} obtained useful data in all X-ray instruments, i.e. the EPIC (European Photon Imaging Camera, 0.2\,--\,12.0~keV)
as well as the RGS (Reflection Grating Spectrometer, 5\,--\,38~\AA\,). The EPIC consists of three detectors, two MOS and one PN,
whereas the PN provides a larger effective area and the MOS a higher spectral resolution. 
For our Altair observation roughly 70\,\% of all detected X-ray photons are from the PN detector. 
All EPIC instruments were operated in `Full Frame' mode with the thick filter to block optical/UV loading, the OM (Optical Monitor) had to be closed due to Altair's optical brightness.
Both observations again consist of two exposures each and include a data gap of a few hours due to telemetry problems;
see also Fig.\,\ref{lc}, which shows the respective X-ray light curves.
The {\it XMM-Newton} observing times of Altair are summarized in Table\,\ref{log}.

\begin{table} [ht!]
\setlength\tabcolsep{5pt}
\begin{center}
\caption{\label{log} X-ray observing log of Altair and observation time of filtered high resolution data.}
\begin{tabular}{lll}
\hline
No. & Time & Obs.time (ks)\\\hline
Obs.01 (ID 0502360101) &17 Oct./18 Oct. 2007 & 61.4 \\
Obs.02 (ID 0502360201) &31 Oct./01 Nov. 2007 & 67.8 \\
\hline
\end{tabular}
\end{center}
\end{table}

{\it XMM-Newton} data analysis was carried out with the Science Analysis System (SAS) version~7.1 \citep{sas} and current calibration files.
Standard selection criteria were applied to the data and periods of enhanced background due to proton flares were discarded from spectral analysis.
EPIC spectra were derived for each instrument and each observations by combining the respective exposures.
To increase the SNR for the analysis of the RGS data, we extracted high resolution spectra only from the PSF core (80\,\%) and
merged all exposures using the tool `{\it rgscombine}'. 
For line fitting purposes we used the CORA program \citep{cora}, using an identical line width and assuming a Lorentzian line shape.
This analysis utilizes total spectra, i.e. we determine one background (+continuum) level in the respective region around each line or group of lines.
Theoretical line emissivities were calculated with the CHIANTI~V\,5.2\,(4.2) code \citep{chi,anti}.

Global spectral analysis was performed with XSPEC~V11.3 \citep{xspec} using multi-temperature models
with variable abundances as calculated with the APEC code \citep{apec}.  
Data of the individual detectors were analyzed simultaneously, but spectra were not co-added.
We find that three temperature components best describe the data
and that no additional absorption column is required.
In our models temperatures, emission measures ($EM=\int n_{e}n_{H}dV$) and abundances of elements with significant features in the measured X-ray spectra
are free parameters, other elemental abundances were set to solar values. All elemental abundances are given relative to solar photospheric values from \citet{grsa}.
Some of the fit parameters are interdependent, especially the absolute values of 
emission measure and abundances of elements with emission lines in the respective temperature range.
Consequently, models with different absolute values of these parameters, but only marginal differences in its statistical
quality, may be applied to describe the data.
All errors derived in spectral fitting are statistical errors given by their 90\,\% confidence range allowing variations of normalizations and respective model parameters.
Additional uncertainties may arise from errors in the atomic data and instrumental calibration, which are not explicitly accounted for.

\section{Results}
\label{ress}
We investigated the X-ray properties of Altair by analyzing its light curves, spectra and emission lines and present our results obtained from different analysis methods in the respective
physical context.

\subsection{Light curves and X-ray variability}

To search for short-term variability on timescales of hours to days, we investigated the temporal behavior of Altair's X-ray brightness
during our 2x2 exposures, that are separated by 14~days and cover in total roughly 35\,h of observation time.
In Fig.\,\ref{lc} we show the count rate in the 0.2\,--\,2.0~keV band as measured by the EPIC, i.e. summed PN and MOS, instrument in time-steps of half an hour.
While clearly no strong flares are present, significant variability of Altair's X-ray brightness at a level of roughly 30\,\% on timescales of hours is seen 
in all individual exposures and over the total observation time. 
This variability could in principle be due to rotational modulation or caused by intrinsic variability of the X-ray emitting features; e.g. microflaring or
emergence/decay of weakly active regions. These possible scenarios can in principle be distinguished by studying the periodicity of the light curve 
or the spectral changes related to the changes in X-ray brightness.
Due to Altair's inclination of about 60\degr\,,only those features being sufficiently close to the surface and being located
at equatorial up to intermediate latitudes would induce a rotational modulation;
in contrast, emission from high and polar latitudes or very extended regions would be always visible. 
Spectral variations are expected to be rather minor in the case of rotational modulation
and for the emergence/decay of weakly active regions, in contrast flaring should be accompanied by a spectral hardening.

\begin{figure}[ht]
\includegraphics[width=92mm]{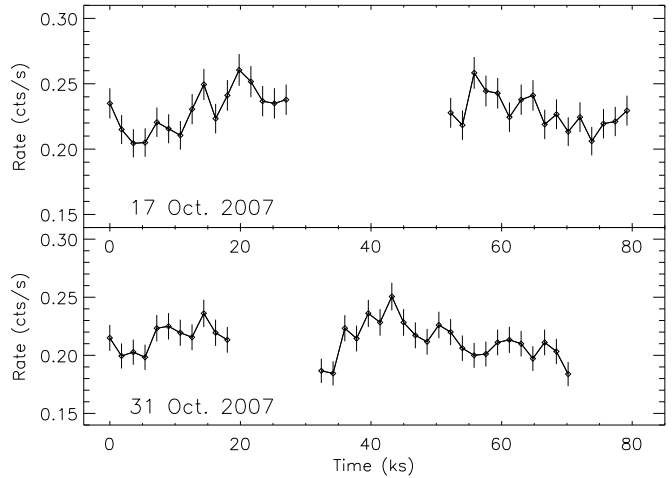}
\caption{\label{lc}Light curves of the Altair observation (EPIC data, 0.2--2.0 keV, 1.8 ks binning).}
\end{figure}

Rotational modulation can be studied with our X-ray data that covers in total three to four stellar rotations. However, 
the uncertainties in the relevant stellar parameters ($Vsini, i, R$) allow a range of rotation periods with values around $9.5\pm 1$~h. 
Further, our data is not continuous, and especially the data gap of two weeks between both observations leads to large phase uncertainties.
Additionally, some restructuring of X-ray emitting surface features might have occurred over times of several days. 
To investigate periodicity of the obtained X-ray light curves, that might be related to rotational modulation,
we tested periods in the range of 7\,--\,13~h for both observations separately, resulting in a minimum of ten time bin pairs per period and observation.
We then calculated the variation of the folded light curve, weighted with the mean deviation, for each period.
As shown in the upper panel of Fig.\,\ref{pha_hr}, a minimum, particularly pronounced for the first observation, is obtained for periods around $10\pm 1$~h.
This indicates, that rotational modulation is indeed present and points to a distribution of active regions that persists at least over a stellar rotation.
The X-ray period is fully consistent with the range of rotation periods derived above and particularly favors values the longer periods.
Neglecting possible differential rotation, that might even be anti-solar, our periods correspondingly suggest values in the lower range for $Vsini$ or in the upper range for the radius.
A period of around 10\,h provides very good self-similarity of the data obtained from the first observation within errors, i.e. over roughly two rotations. 
Likewise it describes the second observation best, however here the scatter is much larger and the X-ray light curve appears rather irregular.
We suspect that short-term variability, emergence and decay of quiescent coronal features or coronal restructuring are very likely responsible for this behavior. 
The here derived X-ray period is in the same range, but slightly larger than a
low-frequency period ($P$\,=\,9.3\,h) found in the pulsation study of \cite{buz05} and the periods derived from modeling of interferometric data,
for example $P$\,=\,9.3\,h \citep{sou05} and $P$\,=\,8.9\,h \citep{pet06}.
If the differences have to be attributed to systematic and measurement errors or indicate different `physical' periods remains uncertain, yet the derived periods of Altair
are close to each other, supporting that rotational modulated X-ray emission is indeed present during the observation.

\begin{figure}[ht]
\includegraphics[width=92mm]{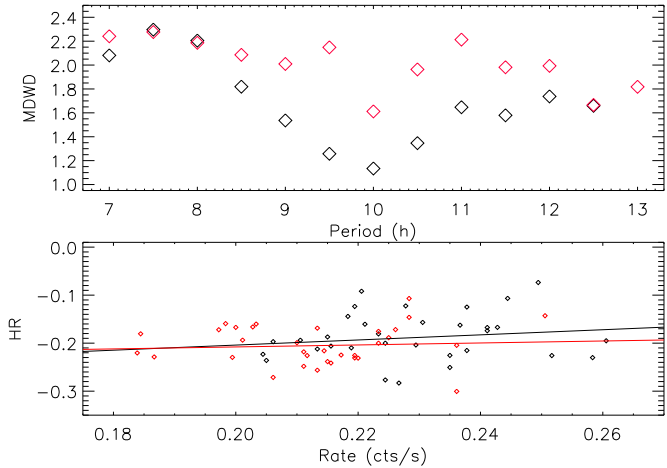}
\caption{\label{pha_hr}X-ray variability of Altair vs. rotation and activity from EPIC data (Obs.\,1: black, Obs.\,2: red). {\it Upper panel}:
Mean deviation weighted differences for the folded light curves vs. period. 
{\it Lower panel}: Hardness ratio vs. X-ray brightness with regression curves.}
\end{figure}

The absence of larger or even moderate flares, which are commonly observed in more active stars, may be a chance effect, but indicates the rareness of such events in stars
with shallow convection zones. The overall X-ray brightness remained fairly constant (mean net count rate of 0.23~cts/s vs. 0.21~cts/s) between both observations, i.e.
over two weeks, corresponding to more than 30~stellar rotations. Further, the X-ray brightness only varies by about 15\,\% around the respective mean value for each observations.
These findings indicate, that the X-ray emitting features are primarily distributed rather uniformly at all stellar longitudes.
Spatially unevenly distributed X-ray emitting regions at low latitudes contribute at most 20\,--\,30\,\% to the total X-ray luminosity.
We note, that some additional contribution from high latitudes or extended regions to the X-ray emission is not in contradiction with the observed light curves.
Overall, the X-ray emitting corona of Altair needs to be rather stable on timescales of several weeks and as shown by the previous X-ray detections, also on timescales of many years.

Intrinsic changes of Altair's surface features might also contribute to the variability of its X-ray brightness.
If this variability is due to weaker unresolved flaring activity as expected for stellar coronae, it should
affect the spectral properties of the observed X-ray emission.
Specifically, a spectral hardening during the X-ray brighter phases should be present;  caused by the higher temperature plasma
in activity related, X-ray emitting features like active regions or small, unresolved flares.
To investigate this possibility, we derive from the PN data for each time-bin a hardness ratio and compare it to the corresponding X-ray brightness.
We use the hardness ratio $HR=(H-S)/(H+S)$ with $S$\,=\,0.2\,--\,0.6~keV and $H$\,=\,0.6\,--\,2.0~keV being the respective photon energy bands.
The spectral range in our soft band mainly covers the X-ray emission from cooler ($\lesssim$~2MK) plasma, 
the harder band traces the hotter plasma.
In the lower panel of Fig.\,\ref{pha_hr} we show the thus obtained spectral hardness vs. X-ray brightness relation for Altair and overlaid linear regression curves for each observation; 
the errors (not shown) on the individual data point are
around 0.01~(Rate) and 0.05~($HR$). The linear regressions are both positive, supporting the picture that
enhanced magnetic activity contributes to the observed variability. Combining the data from both observations, we derive slope of $0.54\pm0.39$. 
However, compared to active stars the slope is rather flat and the scatter is also quite large. 
While extreme spectral variations were not expected given the observed moderate variability, another cause of variability
that does not produce significant spectral changes needs to be present. 

Altogether, our findings confirm the presence of a corona due to magnetic activity.
It is generated by a dynamo that supposable operates in the thin outer convective layer of Altair, predominantly at equatorial up to intermediate latitudes.
Altair's corona is overall stable, whereas rotational modulation, transient coronal features as well as intrinsic variability due to magnetic activity 
contribute to the moderate changes of its X-ray brightness.

\subsection{Global spectral analysis}

To study the coronal plasma properties of Altair and its variability, we performed a global spectral analysis, 
i.e., we modelled the full spectrum of each detector in the spectral range where sufficient source signal is present.
We first investigated the spectra from the two observations separately and found only marginal differences. 
As an example we show in Fig.\,\ref{pnspec} the two PN spectra,
which are obviously very similar. Therefore we determined the global X-ray properties of Altair from all exposures combined and
applied one spectral model to describe all RGS+MOS spectra (the RGS spectrum is shown in Fig.\,\ref{rgs}). 
We examined various multi-temperature models and found that a three temperature model is most suitable to describe the spectra.
We also fitted the high resolution data alone, especially to check the elemental abundances. In this procedure Mg and Si that were 
taken from the combined RGS+MOS fit since no significant features below 10\,\AA\,, that is where these elements produce strong X-ray lines, are present in the RGS spectrum.
A two-temperature model is sufficient to describe the RGS data alone and we find overall consistent results with the combined fit. 
The coronal properties derived from both fitting procedures, i.e. abundances relative to solar values, temperatures and emission measures
as well as the corresponding X-ray luminosity, are given in Table~\ref{par}.

To determine Altair's X-ray activity level, that is described by the $L_{\rm X}/L_{\rm bol}$ ratio,
we additionally calculated the bolometric luminosity from its visual magnitude.
We find $L_{\rm bol}= 3.8 \times 10^{34}$~erg/s, thus Altair is roughly ten times brighter than the Sun.
This value indicates that Altair is slightly evolved, but closer to the main sequence than to the subgiant luminosity class, 
consistent with the isochrone based age estimation of about 1.2\,Gyr \citep{lac99}.

\begin{figure}[ht]
\includegraphics[width=50mm,angle=-90]{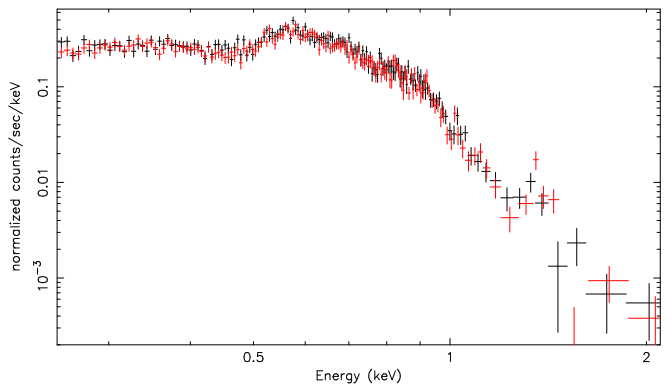}
\caption{\label{pnspec}EPIC PN spectra of the two observations. The spectra are very similar and are dominated by rather cool X-ray emitting plasma}
\end{figure}

\begin{table} [ht!]
\begin{center}
\caption{\label{par} Spectral properties of Altair derived from the XMM-Newton spectra, $L_{\rm X}$ is given in the 0.2\,--\,2.0~keV  band.}
\begin{tabular}{lrrc}\hline\hline
 & MOS/RGS & RGS only & Unit\\\hline
C & 1.20$^{+ 0.69}_{- 0.36}$ & 1.14$^{+ 1.04}_{- 0.40}$& --\\
N &1.13$^{+ 0.62}_{- 0.33}$ & 0.79$^{+ 0.77}_{- 0.34}$& -- \\
O & 0.60$^{+ 0.26}_{- 0.07}$ & 0.76$^{+ 0.59}_{- 0.24}$& --\\
Ne &0.67$^{+ 0.32}_{- 0.17}$ & 0.79$^{+ 0.75}_{- 0.33}$& -- \\
Mg &1.01$^{+ 0.62}_{- 0.39}$ & 1.01 & --\\
Si & 1.56$^{+ 1.84}_{- 1.02}$  & 1.56& --\\
Fe &1.41$^{+ 0.59}_{- 0.23}$ & 1.60$^{+ 0.53}_{- 0.49}$& -- \\
T1& 1.48$^{+ 0.13}_{- 0.12}$& 1.49$^{+ 0.25}_{- 0.17}$& MK\\
EM1 & 4.59$^{+ 0.68}_{- 0.68}$& 5.26$^{+ 1.08}_{- 0.95}$ &$10^{49}$cm$^{-3}$\\
T2& 2.78$^{+ 0.17}_{- 0.12}$ & 2.80$^{+ 0.27}_{- 0.13}$& MK\\
EM2 &5.36$^{+ 0.43}_{- 1.61}$ & 4.95$^{+ 0.83}_{- 0.88}$& $10^{49}$cm$^{-3}$\\
T3& 6.54$^{+ 1.55}_{- 0.96}$ & -- & MK\\
EM3 &0.23$^{+ 0.20}_{- 0.17}$& --& $10^{49}$cm$^{-3}$\\\hline
$\chi^2_{red}${\tiny(d.o.f.)}& 1.12(535)& 1.03 (295) &  \\\hline
L$_{\rm X}$& 1.43& 1.49  & $10^{27}$\,erg/s\\\hline
\end{tabular}
\end{center}
\end{table}

The X-ray luminosity is given in the 0.2\,--\,2.0~keV band, where the EPIC instruments provided useful data. When adopting for comparison the
ROSAT band (0.1\,--\,2.4~keV), the corresponding flux is roughly 20\,\% higher, i.e. we obtain $L_{\rm X}=1.7\times 10^{27}$~erg/s. 
As a crosscheck we also derived X-ray fluxes for each detector independently and find that they overall agree among each other.
Discrepancies on $L_{\rm X}$ are at the 10\,\% level for the presented Altair results, with
the PN detector predicting a slightly higher flux compared to RGS and MOS, especially at energies below 0.5\,keV. 
The very low activity level of $\log L_{\rm X}/L_{\rm bol}= -7.4$, a value that is several ten thousand times below those of active low-mass stars,
confirms that Altair to one of the most inactive stars observed in X-rays. 
Thus magnetic breaking is expected to be very weak. The low X-ray luminosity is related to the low emission measure of $\log EM = 50$~cm$^{-3}$ and
adopting a mean radius of 1.8~R$_{\sun}$, we derive an average surface X-ray flux of $\log F_{\rm X}= 3.9$~erg\,cm$^{-2}$\,s$^{-1}$; similar to values found for solar coronal holes.

Comparing our results with literature data from {\it Einstein}, that measured $\log L_{\rm X}= 27.14$~erg/s in the 0.15\,--\,4.0~keV band \citep{schmitt85}, 
and ROSAT with $\log L_{\rm X}= 27.42$~erg/s \citep{schmitt97}, the {\it XMM-Newton} measurements are intermediate of these values. 
The differences in X-ray luminosity between the observations performed by the three X-ray missions are around 50\,\%
and only slightly larger than the variations already seen within the {\it XMM-Newton} observational campaign. 
It remains uncertain, if cross-calibration issues contribute to the observed differences in X-ray brightness, but the comparison of long-term X-ray data clearly
shows that the corona of Altair is stable and a fairly constant X-ray emitter over several decades.

Altair's corona has an emission measure distribution (EMD) that is dominated by plasma at temperatures in the range of 1\,--\,4~MK,
additionally a weak hotter component with a temperature of 6\,--\,7~MK seems to contribute at a few percent level.
The derived EMD corresponds to an average temperature of about 2.3\,MK, rather typical for weakly active stars and similar to those of the quiescent Sun.
This suggests, that large active regions or a pronounced, but unresolved flaring component are 
virtually absent on the surface of Altair and only minor, moderately active regions somewhat contribute to the X-ray emission.
Rather, its corona is largely dominated by quiescent and weakly active regions or open magnetic field structures.

From our analysis of elemental abundances we find no strong overall metal enhancement or depletion compared to solar values. 
When considering the better determined elements, we find indications for an enhancement of Fe in the corona, especially when compared to O and Ne.
This trend is independent of the used dataset, i.e., it is present in the MOS as well as in the high resolution RGS data.
At optical wavelengths spectroscopic measurements indicated, that the photospheric abundances of Altair differ from the solar values by up to a factor of two, 
with an enhancement of oxygen and a depletion of iron \citep{ers02}. 
This photospheric trend is opposite to the one we have determined for the coronal composition and combined it 
would imply an iron enhancement by a factor of nearly three, while oxygen would be around their photospheric values.
It thus resembles the solar FIP (First Ionization Potential) effect, which, possibly in some modified form, appears also to be present in the corona of Altair that
shows an enhancement of low and medium FIP elements (Fe, Si, C) and nearly photospheric values for high FIP elements (O, Ne).
However, the observed trend is not very strict and larger uncertainties on the abundances for several elements are present. 
The FIP dependence of the coronal abundances is shown in Fig.\ref{fips},
where we plot the coronal to photospheric abundance ratio for the respectively available elements from different abundance determinations. Specifically, we compare
our coronal abundances to photospheric values of the Sun and to those of Altair as determined by \cite{ers02}.

\begin{figure}[ht]
\includegraphics[width=90mm]{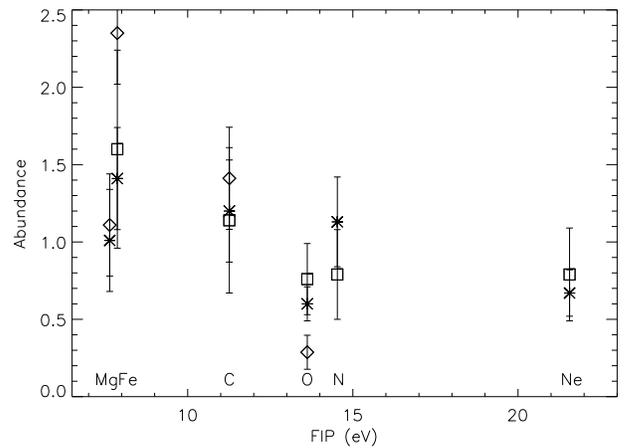}
\caption{\label{fips}Coronal vs. photospheric abundance ratios: asterisks (MOS/RGS vs. solar), squares (RGS vs. solar),  diamonds (MOS/RGS vs. altair).}
\end{figure}

To investigate possible spectral changes that might accompany the X-ray brightness variations, we define two activity levels for Altair, a `low' and a `high' state
based on the EPIC count rate (high $>$~0.22~cts/s, low $<$~0.22~cts/s)
and remodeled the emission measure distribution for both activity states. While the X-ray luminosity and correspondingly the summed EMD is
about 20\,\% higher for the high state, the average coronal temperature is only marginally higher by $\le$~0.1\,MK and within errors both datasets can be described by the same
model with only a renormalization being required. The joint fit of both EPIC data sets results in X-ray luminosities (0.2\,--\,2.0~keV) 
of $L_{\rm X}=1.7~(1.4)\times 10^{27}$~erg/s in the high and low state respectively.
Inspecting the changes in the EMD, we find the largest changes with a factor of roughly two in the hottest component. 
This supports the presence of active regions that induce enhanced variability at higher temperatures.
However, the contribution of the hot plasma to the total emission measure is even in the high state very minor
and changes in the relative contributions of the cool and medium temperature plasma, which overall cancel out, dominate the averaged coronal plasma properties.
This finding is consistent with the rather weak correlation between X-ray brightness and spectral hardness, showing that micro-flaring only plays a minor role in the observed
X-ray variability of Altair.

\subsection{Emission line analysis}

In Fig.\,\ref{rgs} we show the flux-converted RGS1+2 spectrum of Altair. 
The signal below 10\,\AA\, is very weak and no emission lines were significantly detected in this spectral region.
Line fluxes were derived from both RGS detectors where data is available, otherwise only one RGS was used.
We fitted nearby lines, e.g. the oxygen and neon triplet,  simultaneously and note
that the investigated lines are free of strong blends when considering 
Altair's global plasma properties, with the exception of the \ion{Ne}{x} line, which is contaminated by roughly 20\,\% from an iron line blend.
In Table\,\ref{res} we give the measured photon fluxes of the detected spectral lines. The sensitivity of the line flux to e.g. temperature, density and elemental abundance
is used as a diagnostic to investigate the properties of the line forming coronal environment independently of the global fitting procedure.

\begin{figure}[ht]
\includegraphics[width=92mm]{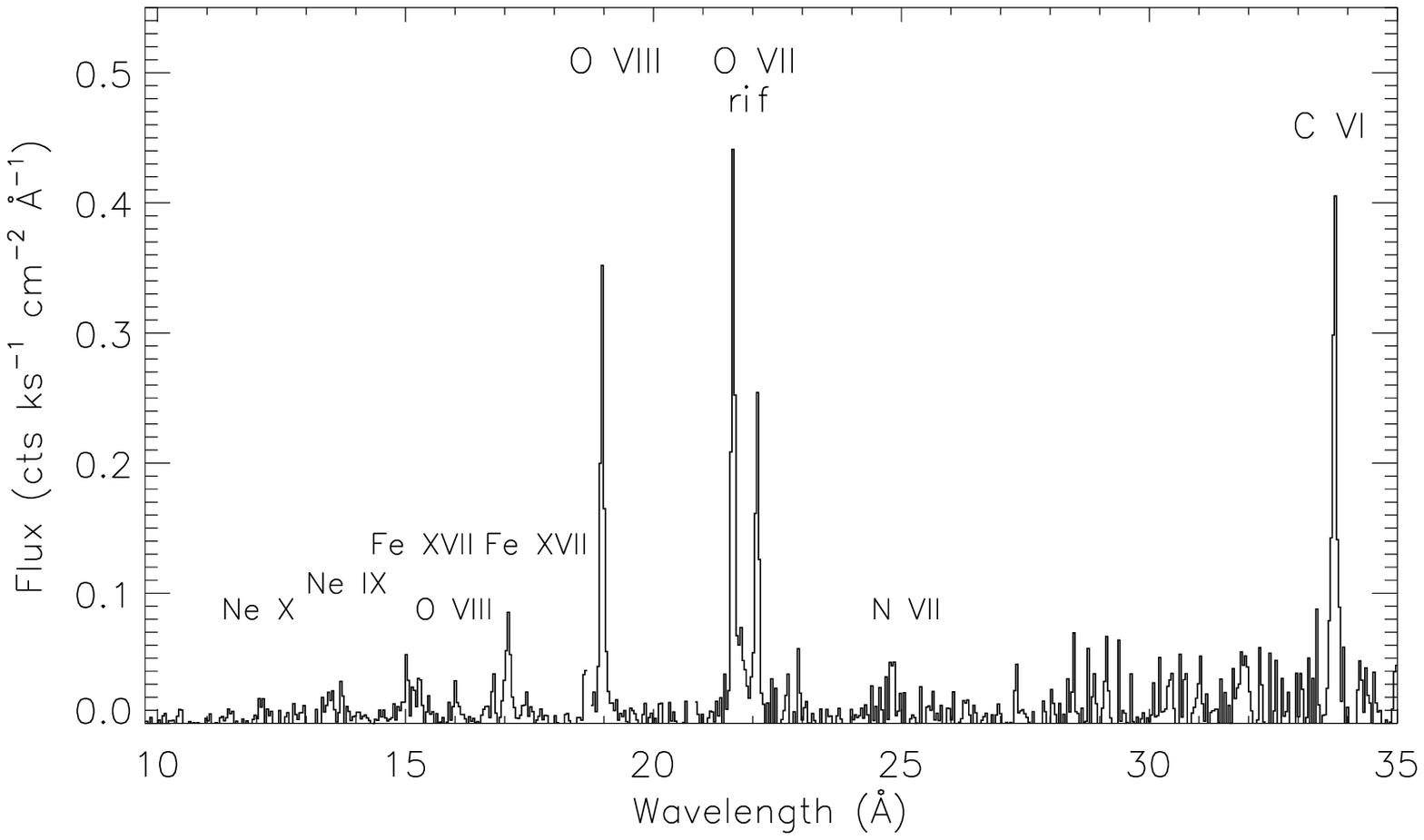}

\vspace*{-4.8cm}
\hspace*{5.5cm}
\includegraphics[width=22mm]{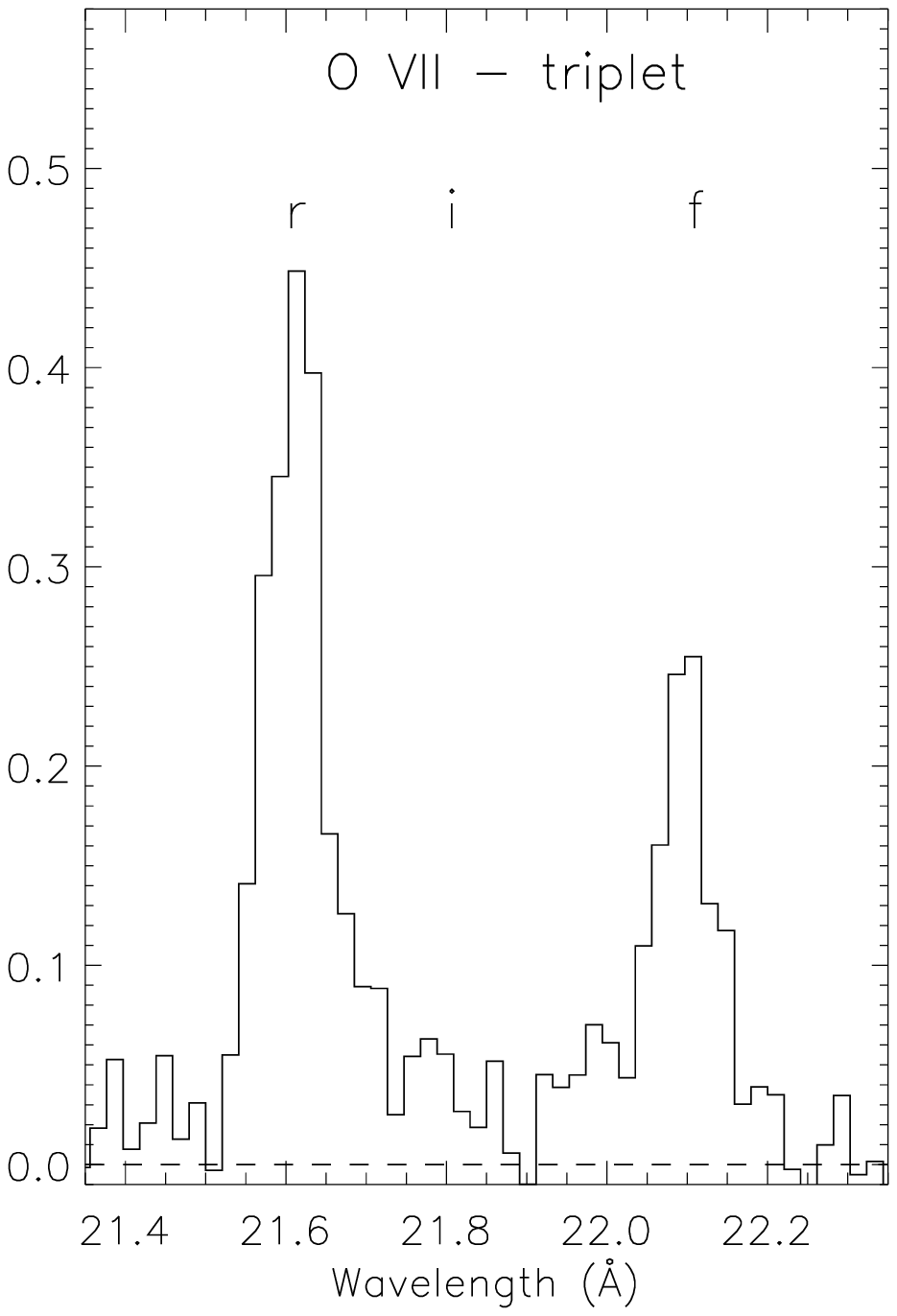}
\vspace*{1.6cm}
\caption{\label{rgs}High resolution X-ray spectrum of Altair with strong emission lines labeled. {\it Inset}:  Zoom in of the \ion{O}{vii} triplet.}
\end{figure}

From the two strongest oxygen lines, \ion{O}{vii}(r) at 21.6\,\AA\, and \ion{O}{viii}\,Ly\,$\alpha$ at 18.97\,\AA\,, we derived an average temperature of the coronal plasma by 
comparing the measured flux ratio to theoretical calculation utilizing the Chianti and APEC database.  
These lines, with peak formation temperatures of around 2\,MK (\ion{O}{vii}) and 3\,MK (\ion{O}{viii}),
are well suited to investigate the temperature of cooler X-ray emitting plasma (1.5\,--\,5.0~MK) as present on Altair,
since their ratio changes by two orders of magnitude over the considered temperature range.
The \ion{O}{viii}\,Ly\,$\alpha$/ \ion{O}{vii}(r) energy flux ratio of 0.86\,$\pm$\,0.10 corresponds to a temperature of 2.3\,--\,2.4~MK, in very good agreement with the
temperature derived from the global fits. 

\begin{table}[!ht]
\caption{\label{res}Line-IDs and measured line fluxes in $10^{-6}$ photons cm$^{-2}$~s$^{-1}$, total RGS data.}
\begin{center}{
\begin{tabular}{lcr}\hline\hline
Ion/Line&Wavelength  (\AA) & Flux \\\hline
\ion{Ne}{x} & 12.14 &2.3$\pm$0.9\\
\ion{Ne}{ix}(r) & 13.45 &4.0$\pm$1.1\\
\ion{Ne}{ix}(i) & 13.55 &1.0$\pm$0.8\\
\ion{Ne}{ix}(f) & 13.70 &3.2$\pm$0.9\\
\ion{Fe}{xvii} & 15.01 &7.0$\pm$1.4\\
\ion{O}{viii} & 16.00 & 4.1$\pm$1.0\\
\ion{Fe}{xvii} & 17.05 &8.0$\pm$2.1\\
\ion{Fe}{xvii} & 17.10 &4.9$\pm$2.0\\
\ion{O}{viii} & 18.97 & 36.6$\pm$3.1\\
\ion{O}{vii}(r) & 21.60 & 48.5$\pm$3.8\\
\ion{O}{vii}(i) & 21.80 & 8.6$\pm$2.2\\
\ion{O}{vii}(f) & 22.10 & 28.7$\pm$3.2\\
\ion{N}{vii} & 24.78 & 9.5$\pm$2.0\\
\ion{C}{vi} & 33.74 & 52.4$\pm$5.1\\\hline
\end{tabular}}
\end{center}
\end{table}

As a further crosscheck of the global fit results and to extend the coronal measurements to the transition region regime,
we used the strongest X-ray lines of each element and ionization state in combination with strong UV lines, taken from \cite{sim97} and \cite{red02}, 
to obtain Altair's EMD over a broader temperature range. 
We included emission lines of \ion{C}{iii}, \ion{C}{vi}, \ion{N}{iii}, \ion{N}{vii}, \ion{O}{vi}, \ion{O}{vii}, \ion{O}{viii}, \ion{Ne}{ix}, \ion{Ne}{x}, \ion{Si}{iii}, \ion{S}{vi} and \ion{Fe}{xvii} with peak
formation temperatures in the range of logT\,=\,4.7\,--\,6.8; albeit with a poorly covered region below logT\,=\,6.0 (between the peak of \ion{O}{vi} and \ion{C}{vi}).
This approach ignores possible variability between the different observations; however, as discussed above, Altair
appears to be a fairly constant source. 
We adopted the fluxes at Earth, neglecting interstellar absorption for Altair and
utilized the 'mcmc\_dem' tool as provided in the PINTofALE~V\,2.6 package \citep{poa} that uses a Markov-Chain Monte-Carlo algorithm to determine the EMD;
here defined as  $DEM(T) = n_{\rm e}^{2} dV/dlogT$. We adopted solar abundances from \cite{grsa} and
compared the theoretical emissivity curves of strong lines with the DEM result. We find the same coronal abundance trends as in the global spectral analysis, i.e.,
an enhancement of C and especially Fe and a depletion of O and Ne. When comparing the results for the X-ray and UV lines, we find that very similar trends are present, e.g.
a low abundance of oxygen is seen in the corona (\ion{O}{vii}, \ion{O}{viii}) and in the transition region (\ion{O}{vi}). 
This suggests, that elemental abundances follow the same trend in both regions and thus fractionation processes would have to occur in lower atmospheric layers. 
We modified the elemental abundances accordingly, however this induced only minor changes to the overall shape of the EMD.
The thus obtained EMD is shown in Fig.\,\ref{dem} and mainly confirms the global fit results;
it exhibits a steep decline in EM at temperatures above 3\,MK (logT\,=\,6.5), but also indicates the presence of a weak component at hotter temperatures.
Altogether, the coronal EM-peak is not very pronounced and resides at rather cool temperatures.

\begin{figure}[ht]
\includegraphics[width=90mm]{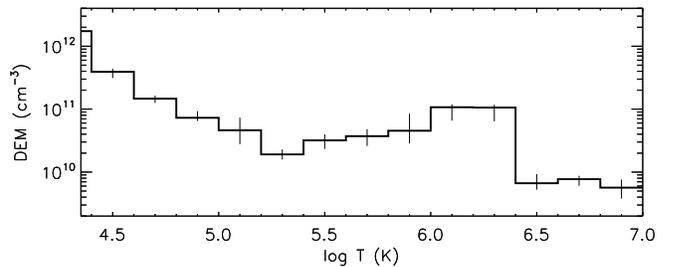}
\caption{\label{dem}EMD of Altair derived from the XMM-Newton, FUSE and HST/GHRS observations.}
\end{figure}

The He-like triplets are used to probe plasma density or the presence of an UV-field \citep{por01}, here we specifically
consider the oxygen triplet. The \ion{O}{vii} $f/i$ ratio is sensitive to the plasma density $n_{\rm e}$ and to the radiation field (UV, 1630\,\AA\,), 
with either high densities or strong UV radiation shifting photons from the forbidden to the intercombination line.
However, when present, both effects combined may likewise account for the measured ratio.
Here we used the relation $f/i=\frac{R_{0}}{1+\phi/\phi_{c}+n_{\rm e}/N_{c}}$ with $f$ and $i$ being the line intensities
in the forbidden and intercombination line, $R_{0}=3.95$ the low density limit of the line ratio,
$N_{c}=3.1\times 10^{10}$ cm$^{-3}$ the critical density and $\phi/\phi_{c}$ the radiation term that describes the photoexcitation rates of the forbidden level \citep{blu72};
values used in the calculations were taken from \cite{pra81}. 

We measure for \ion{O}{vii} a ratio of $f/i = 3.3 \pm 0.9$, slightly below but also consistent with the low density limit or the absence of a radiation field.
Since both effects cannot be disentangled with our data, we discuss in the following the limiting cases; either the ratio is only influenced by the density or by the UV-field.
Adopting the measured $f/i$ ratio (one sigma upper limits in brackets), we derive in the radiation free scenario a plasma density of $\log n_{\rm e}\approx 9.5~(10.2)$~cm$^{-3}$, 
in a low density environment the radiation term becomes $\phi/\phi_{c}=0.2$~(0.65). 
In general, the radiation term depends on the effective temperature of the star, but also on the distance to the stellar
surface. The dilution of the radiation field is described by the factor $W=0.5\,(1-(1-(r_{*}/r)^{2})^{\frac{1}{2}})$, with $r$ being the distance from the center of the star and 
$r_{*}$  the stellar radius \citep{mewe78}. Since Altair's surface temperature also depends on latitude \citep[for surface temperature maps see][]{mon07}, 
we can further constrain the possible emission regions. For a conservative estimate we adopt in the following the upper limit of the radiation term 
and denote the maximum effective temperature for a given distance to the star that is consistent with our value as $T_{\rm max}$.
If the X-ray emitting regions are close to the stellar surface ($r \lesssim$\,1.05~R$_{*}$), we derive $T_{\rm max}=7400$\,K,
i.e. the emission has to originate from low or intermediate latitudes.
In contrast, polar X-ray emission ($T_{\rm eff} > 8000$\,K) would require the emitting location to be at distances $>$\,1.5\,R$_{*}$. 
This would then require strong magnetic fields to confine the plasma, however the absence of strong flaring and of a major hot plasma component as 
well as the tight upper limit on a magnetic field in an extended, e.g. dipole-like structure does not favor this scenario.
We note, that a smaller radiation term would set even tighter constrains, thus it requires the emitting region to be located at cooler stellar latitudes or further away from the star.

Using the above derived results, we are able to estimate the size of Altair's X-ray corona by
adopting the formulae as given in \cite{ness04}, an approach that is based on scaling laws derived for solar coronal loops.
As an input it requires the plasma temperature and its density and relates them to the size of the typical X-ray emitting coronal structures. 
The height of the loop structures then defines the available coronal volume,
the emitting coronal volume is calculated from the emission measure and the density, and the ratio of the two is denoted as coronal filling factor.
Taking our results for Altair, i.e. $T$\,=\,2.4\,MK, $\log EM$\,=\,50~cm$^{-3}$, $\log n_{\rm e}$\,=\,9.5~cm$^{-3}$ and $R$\,=\,1.8~R$_{\sun}$, 
we obtain a loop (half-)length for the \ion{O}{vii} emitting structures of about $L=2 \times10^{9}$~cm (0.02 R$_{*}$) and a coronal filling factor of 0.03.
Both parameters scale with the inverse of the plasma density, that has the largest uncertainty of the parameters involved and may vary by a factor of up to five.
Therefore, the given values have to be taken as a rough estimate
and they are formally a lower limit when the possible low density limit and/or presence of the radiation term are considered.
However, significantly lower densities require much larger coronal structures and also imply a larger filling factor, raising again the question of plasma confinement.
Therefore, the loop structure analysis as well as the study of possible plasma locations under consideration of the UV-field both lead to very similar conclusions.
We have also investigated the $f/i$ ratio from the \ion{Ne}{ix} triplet that traces hotter plasma; the derived results are consistent with those from the oxygen triplet,
however they do not add further information.

The analysis of emission lines reinforces the picture derived from the study of the light curves and the global spectra.
We find that in the most likely scenario the coronal structures on Altair are quite compact compared to the stellar dimensions 
and their typical size is about one to a few percent of the stellar radius.
The coronal plasma is cool, at relatively low densities and fills up to a few percent of the available coronal volume.
As a whole, Altair is rather sparsely covered with X-ray emitting structures, with
the X-ray emission mainly originating from the cooler surface areas, i.e. equatorial regions or intermediate latitudes.

\subsubsection{Coronal neon and oxygen abundances}

Over the last years there has been some debate on the true neon content of the Sun and other stars \citep[see e.g.][and therein]{rob08}.
The controversy arose due to a disagreement between helioseismology and a downward revision of solar abundances.
We previously studied Ne/O ratios in a sample of low and moderately active stars of later spectral type (mid-F to mid-K).
The basic outcome was that, first the coronal Ne/O ratio is activity dependent with more active stars
showing a higher Ne/O ratio, and that second stars with solar-like activity level show roughly solar-like Ne/O abundance ratios.
Since Altair also has a very low activity level, our data is well suited to extend the stellar sample towards hotter photospheric temperatures.

\begin{figure}[ht]
\includegraphics[width=90mm]{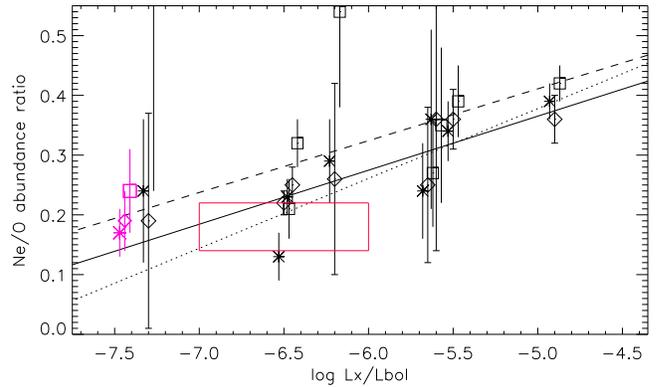}
\caption{\label{neo}Coronal Ne/O ratio of Altair (purple) combined with results for our sample stars of later spectral type (diamonds (solid): global, 
asterisks (dotted): line-ratio 1, squares (dashed): line-ratio 2)
and parameter range of the `classical' Sun, shown as red box \citep[see][]{rob08}. }
\end{figure}

Beside the global fit result derived above, we determined the Ne/O ratio by utilizing linear combinations of
emission lines from neon and oxygen that should provide a roughly EMD independent abundance ratio.  
We used two different line ratios as in our previous work to derive the Ne/O ratio, i.e.
the energy flux ratio (1): \ion{O}{viii} vs. $\ion{Ne}{ix}+0.15\times\ion{Ne}{x}$
and the photon flux ratio (2): $0.67\times\ion{O}{viii}-0.17\times\ion{O}{vii}$ vs. $\ion{Ne}{ix}+0.02\times\ion{Ne}{x}$. 
We note that for the coronal properties of Altair the ratios from the two linear combinations of emission lines rather supply lower and upper boundary values for the true Ne/O ratio;
a full discussion of the systematic errors of these methods can also be found in \cite{rob08}.

We thus derived the Ne/O ratios for Altair and compared these to the solar abundance ratios.
\cite{grsa} give a solar ratio of Ne/O\,=\,0.18\,$\pm 0.04$, we derived for Altair from the global fit Ne/O\,=\,0.20\,$\pm 0.05$ and with RGS data only Ne/O\,=\,0.19\,$\pm 0.07$, from the two
emission line ratios we derived Ne/O\,=\,0.17\,$\pm 0.04$~(1) and Ne/O\,=\,0.24\,$\pm 0.07$~(2). 
Altogether this gives a mean coronal ratio for Altair of Ne/O\,=\,0.20\,$\pm 0.03$, slightly above but fully consistent
with the `classical' solar one. The Altair result and the ratios from our sample stars of later spectral type are
shown in Fig.\,\ref{neo}, the respective linear regressions are overlaid.
Altair's low Ne/O ratio fits well in the trend derived from the cooler stars, suggesting similar chemical fractionation processes and coronal abundance patterns.

\subsection{The activity-rotation correlation}

The onset of magnetic activity in late A-type stars as well as the regime of very inactive stars are only poorly studied at X-ray wavelengths.
Its spectral type and very low activity level place Altair into both categories at once, making suitable objects for comparison rare. 
Nevertheless, given its repeated X-ray detections over several decades, 
a very inefficient but quite stable dynamo mechanism needs to operate somewhere in the thin outer convective layer of this fast rotating star.

As a tentative approach for comparison with late-type stars we adopt a solar-like dynamo to generate the observed X-ray emission from Altair and 
discuss its possible implications. When assuming an $\alpha\,\Omega$ dynamo,
the correlation of magnetic activity with rotation or more precisely dynamo efficiency in the non-saturated regime is usually described by the Rossby number, 
i.e. Ro\,=\,$P/\tau_{c}$ with $P$ being the rotation period and $\tau_{c}$ the convective turnover time. Its validity has been shown over a broad range of stellar activity
for chromospheric \ion{Ca}{ii} emission \citep{noy84} and coronal X-ray emission \citep[e.g.][]{ran00}. However, in both cases the region of weakly active stars with log~Ro\,$>0.5$ 
is nearly unexplored and further the samples are dominated by less massive ($\lesssim 1{\rm M}_{\sun}$) stars.
Adopting the relation $L_{\rm X}/L_{\rm bol} \propto {\rm Ro}^{-\alpha}$ with $\alpha \approx 2$ that is valid for large Rossby numbers (log~Ro\,$\gtrsim -0.8$),
we can derive the dynamo efficiency for Altair and compare the expected with the observed activity level.
To obtain a rough estimate of Altair's convective turnover time, we adopt an empirical formula, although derived from less massive late-F to late-K  stars,
that relates $\tau_{c}$ with B-V color as given in \cite{noy84}.
For Altair's B-V\,=\,0.22, corresponding to $T_{\rm eff}=$\,7650\,K, we obtain $\tau_{c}= 1.3$\,h 
and when adopting a period of 9.6\,h we finally get log~Ro\,=\,0.88. Inspecting the correlation shown in \cite{ran00},
derived from a stellar sample that includes many ROSAT field stars, we obtain for large Ro the empirical relation $\log L_{\rm X}/L_{\rm bol} = -4.9 -2.1\times {\rm log~Ro}$.
With the values for Altair we then expect $\log L_{\rm X}/L_{\rm bol}\approx -6.8$, 
a value of the same order of magnitude but roughly a factor four larger than the observed value of $\approx -7.4$. 
Given the extrapolation and Altair's latitude dependent temperature and radius, this rough agreement is remarkable, but could be a chance effect.
While a larger amount of the hot plasma might be lost as coronal winds from open magnetic structures,
a modified dynamo model seems to be physically more adequate. This model has to consider the particular properties of Altair,
specifically its distorted structure and surface, here accounted for by the latitude dependent photospheric temperature. 
We note that the similar A7 star Alderamin ($\alpha$\,Cep) has a RASS X-ray luminosity comparable to that of Altair, indicating that the coronal properties of Altair are rather characteristic for
this type of star. 

Instead of a globally operating solar-like dynamo, we therefore additionally investigated a localized version to describe the magnetic field generation in Altair. 
\cite{dob89} and \cite{ran96} find that the activity-rotation relation holds up to B-V\,=\,0.3, i.e. the regime of early F-type stars which have
effective temperatures similar to the temperatures that exist in the equatorial region of Altair. Repeating the above calculation for B-V\,=\,0.3 ($\approx$\,7000\,--\,7200~K) leads to 
an expected activity level of $\log L_{\rm X}/L_{\rm bol}\approx -5.3$. Consequently, when generating X-rays with a similar efficiency,  
only a few percent of Altair's surface would easily account for the observed X-ray emission. 
The surface area exhibiting these photospheric temperatures is supposably much larger, pointing
to a less efficient dynamo mechanism in the localized version. In this scenario, magnetic activity occurs mainly at the cooler, lower latitudes, consistent with
the X-ray emitting regions found in our analysis.

Likely in the most massive stars with thin outer convection zone the $\alpha - \Omega$ effect is only locally operating or
not the dominant dynamo mechanism at all and hence a global Rossby number is not the adequate property to describe their activity level. 
Considering additionally the latitude dependence of radius and temperature that result from the fast rotation, differing dynamo characteristics seem inevitable.
Comparing the X-ray results with those obtained at other wavelengths, it appears that
the trend of diminished activity in the hotter A-type stars is much more pronounced in the X-ray regime when compared to chromospheric measurements \citep{sim97}.
A similar trend may also be present, when comparing transition region lines to chromospheric emission, but adequate data becomes quite sparse in this regime.
This would favor an overall diminishment of hot plasma in the outer stellar atmosphere of A-type stars in general, instead of a shift of its temperature.

On the other hand, the very low activity level of Altair might indicate a turnover in the activity-rotation relation for weakly active stars, i.e. around $\log L_{\rm X}/L_{\rm bol}= -6$.
Other inactive stars of later spectral type, e.g. Procyon~(F5) with $\log L_{\rm X}/L_{\rm bol}= -6.5$, log~Ro\,=\,0.9/1.2 (B-V\,=\,0.4, P\,=\,10/20\,d) and $\alpha$\,Cen~A~(G2) 
with $\log L_{\rm X}/L_{\rm bol}= -7$, log~Ro=0.3 (B-V\,=\,0.7, P\,=\,29\,d) show activity levels that are likewise by up to an order of magnitude below the expected value from the above correlation.
Altogether, the observed X-ray activity levels in very inactive stars exhibit a quite large spread, but are rather low when compared to expected values based on their Rossby number. 
Whether the spread or a possible turnover of the correlation between activity and rotation is related to mass or other stellar properties
cannot be investigated here due to very limited statistics. However, the available data suggests that weakly active stars might be overall
less active than expected from a simple down-scaling of their more active counterparts. 

An X-ray activity-rotation study that includes many ROSAT field stars, confirms
the above described correlation by using an empirical X-ray Rossby number and the activity level, albeit only stars with B-V\,$>$\,0.5 were considered \citep{piz03}.
This study also suggests a possible mass dependence of the maximum activity level; especially towards more massive stars a decrease of the maximum activity level is indicated.
However, within errors a saturation level of $\log L_{\rm X}/L_{\rm bol} = -3.2$ is acceptable for all given mass ranges (0.2\,M$_{\sun}<$\,M\,$<$\,1.2\,M$_{\sun}$)
and additionally a large scatter is present at low activity levels for the most massive sample stars.
Extrapolating their (almost) mass independent scaling law for the saturation limit leads to the conclusion, that Altair's activity is actually around the X-ray saturation limit
for stars of spectral type A7, although it is roughly four orders of magnitude below the corresponding value for late-type stars.

Altair is rotating at a considerable fraction ($\gtrsim 60\%$) of its break-up speed, thus a significant higher X-ray activity level is ruled out for an $\alpha\,\Omega$ dynamo,
as well as for any other dynamo mechanism, whose efficiency roughly scales linearly with rotation.
An activity level of $\log L_{\rm X}/L_{\rm bol} \approx  -3$, the typical saturation value for magnetic activity in less massive stars with well-developed convection zones, 
is by far out of reach for Altair.
Therefore the X-ray observation of Altair implies that the X-ray activity saturation level for A7 stars (M\,$\approx$\,2\,M$_{\sun}$) 
is by about four orders of magnitude diminished, when compared to solar-like or late-type stars.

\section{Discussion}
\label{dis}
Combining our results obtained from the X-ray data of the A7 star Altair, we are able to derive a picture of the basic structure of its corona, it's X-ray properties and
the underlying dynamo mechanism. The results from the X-ray observation are compared to findings for the other outer atmospheric layers.

Altair is a weakly active star with a corresponding very low X-ray activity level.
The observed X-ray emission is dominated by relatively cool plasma; larger flares and hot active regions are virtually absent on Altair,
however moderate activity seems to contribute at a minor level. Its X-ray brightness is fairly constant on timescales of days up to decades. Rotational modulation
and emergence/decay of coronal structures mainly account for the moderate variations in X-ray brightness, which is not accompanied by significant spectral changes.
The spectral properties are reminiscent of those from
small-scale magnetic structures or quiescent regions on the Sun and other weakly active stars. The coronal plasma is at low density and its chemical composition
shows a FIP-effect like abundance pattern with a Ne/O ratio around solar value. Therefore, despite being more massive and having higher photospheric temperatures,
Altair's coronal X-ray properties are very similar to those of similarly active stars of later spectral type.

We find that  equatorial regions or low to intermediate latitudes are the most X-ray active areas of Altair's surface, 
i.e. the same regions that exhibit the lowest surface temperatures \citep{mon07}.
This is in contrast to fast rotating, but less massive stars like the K2 dwarf BO~Mic, where magnetically active regions
are predominantly found at high latitudes and polar regions \citep{wol08}.  
Magnetic activity, that is mainly located at low latitude regions
is independently suggested by a modulation of the X-ray light curve with the rotation period, as well as from the study of possible plasma locations derived from emission line ratios.
The analysis of the size of the coronal structures supports this finding and indicates, that Altair's corona consists of rather small coronal loops and has an overall small filling factor.
In this scenario, the X-ray emission is related to magnetic activity, that is predominantly generated in the equatorial bulge and neighboring regions.

The equatorial regions of Altair's surface have photospheric temperatures of an early F-type star, and such stars are known to exhibit X-ray emission. 
Very thin plasma residing in very large coronal structures, that contributes to the observed X-ray emission, is in principle not contradicting our data, 
but its confinement and the stability of such structures is questionable given Altair's fast rotation and absence/weakness of large scale magnetic fields \citep{lan82}.
A compact emission region located close to the surface of Altair is also supported by emission lines originating in the chromosphere and transition region.
The \ion{O}{vi} lines at 1032\,\AA\, and 1038\,\AA\,, tracing plasma roughly at $3\times 10^{5}$~K that is located in the transition region, i.e. from the atmospheric layer below the corona,
are also apparently formed close to the stellar surface.
Their broad, double-horned profile is well fitted by optically thin emission originating on a star with $Vsini=210$~km/s, that exhibits either strong limb brightening or emitting regions
preferentially located towards the equator \citep{red02}. A chromosphere covering mainly equatorial regions was also suggested as a possible scenario by \cite{fer95}, 
deduced from the modelling of the \ion{H}{i} Ly\,$\alpha$ line profiles.

While not much is known about the interior structure of Altair, the star is clearly oblate and its surface is highly distorted. 
Therefore, most possibly also the outer part of the interior is not uniform and it seems reasonable to associate the development of convective cells
with the more cooler surface areas near the equator. In this scenario, localized dynamo processes significantly contribute to the observed magnetic activity. 
This would naturally explain low latitude emission and, since
only a fraction of Altair's surface is magnetically active, could be a main cause for its very low activity level.
Additionally, the locally operating dynamo appears to be less efficient in generating X-ray flux per surface area, when compared the global, solar-like dynamo for the same
effective temperature and rotation period.
However, surprisingly low activity levels also seem to be present in other weakly active stars of later spectral type.
Altogether, the cause for the diminished stellar activity levels remains uncertain and quite possible it differs for stars with shallow convection zones and those with slow rotation.
On the other hand, the weak X-ray activity level of Altair seems to be also the highest possible activity level, since a significant spin-up would disrupt the star.
Consequently, the fast rotator Altair is likely already around the saturation limit for magnetic X-ray activity when considering its dynamo efficiency or rotation period.
This implies, that the X-ray saturation level in stars with very shallow convection zones around spectral type A7 is about four orders of magnitude below the
saturation level for later-type stars.
No matter what are the details of the generating mechanism that is underlying Altair's magnetic activity, its corona appears to be fairly stable on timescales from days to decades.

It is highly desirable to study of other, ideally also more slowly rotating mid to late A-type stars, to derive a more complete picture of magnetic and X-ray activity related
phenomena at the onset of the development of outer convection zones. The here presented X-ray observation of the A7 star Altair leads, 
together with chromospheric and transition region measurements,
to an overall comprehensive and consistent picture of the magnetic activity for this fast rotating, distorted star.

\section{Conclusions}
\label{con}

   \begin{enumerate}
      \item  The A7 star Altair has an X-ray luminosity of $L_{\rm X}=1.4\times 10^{27}$~erg/s in the 0.2\,--\,2.0~keV band,
resulting in a very low activity level of $\log L_{\rm X}/L_{\rm bol}= -7.4$.
Its X-ray brightness exhibits rather smooth variability at the 30\,\% level, 
mainly due to rotational modulation and occurrence or restructuring of coronal features. Some minor activity is present, in contrast larger flares or significant spectral changes that
accompany the X-ray brightness variations were not found during the {\it XMM-Newton} observation.

\item Altair's X-ray properties are very similar to those of other weakly active stars like the quiescent Sun.
Its corona is dominated by cool plasma (1\,--\,4~MK), leading to an average plasma temperature of less than 2.5\,MK.
The coronal plasma is at low density, elemental abundances  show a solar-like FIP effect and Ne/O ratio.
The hotter photosphere of Altair does not result in major differences of the X-ray emitting plasma, when compared to the Sun and other weakly active stars.

\item With its fast rotation and thin outer convective layer, Altair provides an overall inefficient, but very stable dynamo mechanism as deduced from its repeated X-ray detection over several decades at an
fairly constant X-ray brightness level. 
X-ray data favor the coronal structures to be compact and to be located predominantly around the equator or at low latitude regions.
Here, where the radius is largest and the effective temperatures of Altair's distorted surface are lowest, the main dynamo action takes place.

\item The X-ray activity saturation level of Altair, and probably around spectral type A7 in general, is reduced by roughly four orders of magnitude 
when compared to typical values for less massive late-type stars.
   \end{enumerate}

\begin{acknowledgements}
This work is based on observations obtained with XMM-Newton, an ESA science
mission with instruments and contributions directly funded by ESA Member
States and the USA (NASA). J.R. acknowledges support from the DLR under 50QR0803.

\end{acknowledgements}

\bibliographystyle{aa}
\bibliography{1348}

\end{document}